\documentstyle{article}
\begin{document}
\vspace{-2cm}
\title{\sc Diffractive Dissociation In The Interacting Gluon Model}
\author{F.O. Dur\~aes$^1$\thanks{e-mail: dunga@uspif.if.usp.br}, \ F.S.
Navarra$^{1,2}$\thanks{e-mail: navarra@uspif.if.usp.br} \ and \ G.
Wilk$^2$\thanks{e-mail: wilk@fuw.edu.pl} \\ 
{\it $^1$Instituto de F\'{\i}sica, Universidade de S\~{a}o Paulo}\\
{\it C.P. 66318,  05389-970 S\~{a}o Paulo, SP, Brazil} \\[0.1cm]
{\it$^2$Soltan Institute for Nuclear Studies, 
Nuclear Theory Department}\\
{\it ul. Ho\.za 69, \ Warsaw, Poland}}
\maketitle
\vspace{1cm}
\begin{abstract}
We have extended the Interacting Gluon Model (IGM) to calculate 
diffractive mass spectra generated in hadronic collisions. We
show that it is possible to treat both diffractive and 
non-diffractive events on the same footing, in terms of 
gluon-gluon collisions. A systematic analysis of available data
is performed. The energy dependence of diffractive mass spectra
is addressed. They show a moderate narrowing at increasing energies. 
Predictions for LHC energies are presented.
\\

PACS number(s): 13.85.Qk, 11.55.Jy\\

\end{abstract}

\newpage
\section{Introduction}

In the last years, diffractive scattering processes have received
increasing attention for several reasons. These processes may, for
example, explain many features of particle production and, in
particular, heavy flavour production \cite{HF} and Centauro events
\cite{CAP}. They are also related to the large rapidity gap physics
and the structure of the Pomeron \cite{POMERON}. In a diffractive
scattering, one of the incoming hadrons emerges from the collision
only slightly deflected and there is a large rapidity gap between it
and the other final state particles resulted from the other excited
hadron. In some models diffraction is due to the Pomeron exchange 
but the exact
nature of the Pomeron in QCD is not elucidated yet. The first test of
a theory (or a model) of diffractive dissociation (DD) is the ability
to properly describe the mass ($M_X$) distribution of diffractive
systems, which has been measured in many experiments \cite{DATA} and
parametrized as $(M_X^2)^{-\alpha}$ with $\alpha \simeq 1$. Data
presented in \cite{DATA} were taken at the Tevatron collider
($\sqrt{s} = 1.8$ TeV). They allow us to make comparison with CERN
\cite{Bozzo} lower energies data and observe the energy dependence of
the mass spectrum. In the old Regge theory, the assumption of Pomeron
dominance implies that the mass spectrum behaves like $1/M^2_X$ and
does not depend on the energy \cite{REGGE} whereas in \cite{DATA} a
slight deviation from this behaviour was reported.\\

In this work we intend to study diffractive mass distributions
using the Interacting Gluon Model (IGM) developed by us recently
\cite{IGM,DNW94,DNW93}. In particular, we are interested in the
energy dependence of these distributions and their connection with
inelasticity distributions \cite{INEL,DNW94}. One advantage of the
IGM is that it was designed in such a way that the energy-momentum
conservation is taken care of before all other dynamical aspects.
This feature makes it very appropriate for the study of energy flow
in high energy hadronic and nuclear reactions \cite{IGM,DNW94,DNW93}
and in cosmic ray studies \cite{INEL}. In particular, as shown in
\cite{DNW94,DNW93}, the IGM was very usefull in analysing data and
making predictions on the behaviour of inelasticities and leading
particle spectra, including leading charm production \cite{DNW96}.
The aim of the present work is to demonstrate that the main
characteristics of the DD processes mentioned above emerge
naturally from the standard IGM, enlarging profoundly its range
of applications.\\
 
In the following section we present for readers convenience the basic
elements of the IGM. It is then applied to DD processes in Section 3 
where we also carefully explain what DD means in terms of the IGM and
how it differs from the conventional Regge Pomeron approach. Section 4 
contains our numerical results and comparison with data and the last
section is devoted to our conclusions.\\

\section{Interacting Gluon Model}
\subsection{General ideas}
The Interacting Gluon Model is based on the following  idea \cite{POK} :
since about half of a hadron momentum is carried by gluons and since
gluons interact more strongly than quarks, during a collision there
is a separation of constituents. Valence quarks tend to be fast forming
leading particles whereas gluons tend to be stopped in the central
rapidity region. This picture is consistent with string formation and
fragmentation as it is formulated in the Lund Model or in the Dual 
Parton Model. These models are based on the concept of string and were
constructed to work at low average transverse momenta. When, in the
late eighties, the energy of hadronic collisions increased by more
than one order of magnitude it became necessary to incorporate the
concept of parton and of hard and semi-hard collisions. The latter
are collisions between partons at a moderate scale 
($ Q^2 \simeq (2\, GeV)^2 $) which however still allows for the use of
perturbative QCD. The scattered partons form the so-called minijets.
At $\sqrt{s}= 540 $ GeV the minijet cross section is already $25 \%$
of the total inelastic cross section. A new generation of models 
appeared using the concept of parton and the QCD parton model formalism
and trying to understand minimum-bias multiparticle physics extending
the parton model to semihard (moderate) energy scales. Among these
models we mention those presented by Gaisser and Stanev \cite{gs},
Sjostrand \cite{sj}, Wang \cite{wang} and Geiger \cite{geiger}. The
IGM belongs to this class of models. In all these models one finds
at a certain point expresssions of the type
\begin{eqnarray}
 \int_{Q^2}^1\!dx_1\, \int_{\frac{Q^2}{x_1}}^1\! dx_2\,
f(x_1,Q^2)\,f(x_2,Q^2)\,\sigma(x_1,x_2,Q^2) 
\end{eqnarray}
where f and $\sigma$ are parton momentum distribution and parton-parton
elementary cross section respectively. $Q^2$ is the scale. Apart from
some ambiguity in choosing the scale, these models have to face the
problem that even at very high energies a significant part of a hadronic
collision occurs at scales lower than the semihard one. At this point
f and $\sigma$ are not very well known. The atitude taken in HIJING
\cite{wang}, in the Parton Cascade Model \cite{geiger} and also in the
IGM is to extrapolate these quantities to lower scales. These 
extrapolations can be continuously improved, especially in view of the
advance of our knowledge on non-perturbative effects. There are, for
example, models for distribution functions
which work at scales as low as $ 0.3\, GeV^2$ \cite{grv}. As for $\sigma$
one can compute non-perturbative effects in the context of an operator
product expansion \cite{nos}. Inspite of these limitations these models
have the advantage of dealing with partons and being thus prepaired to
incorporate perturbative QCD in a natural way. This is welcome since 
perturbative processes are expected to be increasingly important at
higher energies.  Compared to the other models mentioned above, the IGM
is simpler because it is designed to study energy flow and makes no
attempt to calculate cross sections or to follow hadronization in great
detail. This simplifies the calculations and avoids time consuming numerical
simulations. The most important aspect of the IGM, shared with those models,
 is the assumption of multiple parton-parton incoherent scattering which is
implicit  in the Poissonian distribution of the number of parton-parton
collisions (which is also used in refs. \cite{sj,gs,wang}) used below. In 
going to lower resolution scales this independent collision approximation
becomes questionable. On the other hand lattice QCD calculations \cite{gia}
in the strong coupling regime indicate that the typical correlation length of
the soft gluon fields is around $0.2$ fm. This is still smaller than the typical
hadronic size. Therefore the independent collision approximation may still be a
reasonable one. From the practical point of view, it was shown in \cite{IGM}
that replacing the Poissonian distribution by a broader one does not affect the
results significantly as long as some mass scale is introduced to cut off the
very low x region.
\subsection{Formulation of the model}

The IGM is based on the assumed dominance 
of hadronic collisions by gluonic interactions \cite{POK} and can be
summarized as follows \cite{IGM} (cf. Fig. 1a):
\begin{itemize}
\item[$(i)$] The two colliding hadrons are represented by valence
quarks carrying their quantum numbers (charges) plus the accompanying
clouds of gluons (which represent also the sea $q\bar{q}$ pairs and
therefore should be regarded as effective ones).
\item[$(ii)$] In the course of a collision gluonic clouds interact
strongly and form a gluonic {\it central fireball} (CF) located in the
central region of the reaction.
\item[$(iii)$] The valence quarks (plus those gluons which did not
interact) get excited and form {\it leading jets} (LJ's) (or {\it beam
jets}) which then populate mainly the fragmentation regions of the
reaction.
\end{itemize}
It should be stressed that the IGM has been 
formulated originally in order to
give the initial conditions for hydrodynamical models by providing in a
dynamical way the so called {\it inelasticity} of the reaction understood
usually as the (invariant) fraction of energy stored in the CF:
\begin{equation}
K\, =\, \sqrt{x\cdot y} \label{eq:K}
\end{equation}
with $x$ and $y$ being the fractions of the initial energy-momenta of
the respective projectiles allocated to the CF. This variable and, in
particular, its energy dependence is of vital importance in cosmic
ray studies as it is the  necessary ingredient allowing to deduce any
elementary information from cosmic ray experiments \cite{INEL}.

According to the IGM, pairs of gluons collide and form ``minifireballs''
(MF's). In collisions at higher scales a minifireball is the same as a
pair of minijets or jets. In the study of energy flow the details of
fragmentation and hadron production are not important. Most of the MF's
will be in the central region and we assume that they coalesce forming
the CF. The probability to form a fireball carrying momentum fractions 
 $x$ and $y$  of two colliding hadrons is defined as the sum over
(an undefined number $n$ of) MF's \cite{IGM}:
\begin{eqnarray}
\chi(x,y) &=& \sum_{\{n_i\}} \delta \left( x-\sum_i n_i x_i \right)
\delta \left( y-\sum_i n_i y_i \right) \prod_{\{n_i\}} P(n_i)
\end{eqnarray}
(all masses and transverse momenta are neglected in what follows).
The distribution of the number of MF's is given by $P(n_i)$ for which
we use Poisson distributions:
\begin{eqnarray}
P(n_i) &=& \frac{(\overline n_i)^{n_i} exp(-\overline n_i)}{n_i\!}
\end{eqnarray}
corresponding to independent production. Expressing the delta functions
via Fourier integrals one can perform all summations, transform certain
summations over $\overline n_i$ into integrals and make the replacement
$\ d \overline n_i / dx dy = \omega(x,y)$ arriving at the 
general formula  \cite{IGM}:
\begin{eqnarray}
\chi(x,y) &=& \frac{\chi_0}{2\pi\sqrt{D_{xy}}}\cdot \label{eq:CHI}\\
&&\cdot \exp \left\{ - \frac{1}{2D_{xy}}\,\left[
  \langle y^2\rangle (x - \langle x\rangle )^2 +
  \langle x^2\rangle (y - \langle y\rangle )^2 -
  2\langle xy\rangle (x - \langle x\rangle )(y - \langle y\rangle )
  \right] \right\} ,\nonumber
\end{eqnarray}
where
\begin{eqnarray}
D_{xy} &=& \langle x^2\rangle \langle y^2\rangle - 
           \langle xy\rangle ^2  \nonumber 
\end{eqnarray}
and
\begin{eqnarray}
\langle x^ny^m\rangle &=& \int_0^1\! dx\,x^n\, \int_0^1\! dy\, 
y^m\, \omega (x,y), \label{eq:defMOM}
\end{eqnarray}
with $\chi_0$ being a normalization factor defined by the condition
that
\begin{eqnarray}
 \int_0^1\!dx\, \int_0^1\! dy\, \chi(x,y) \Theta(xy - K_{min}^2) = 1
\end{eqnarray}
with $K_{min} = \frac{m_0}{\sqrt{s}}$ being the minimal inelasticity
defined by the mass $m_0$ of the lightest possible CF. 
The, so
called, spectral function $\omega(x,y)$ contains all the dynamical
input of the IGM in the general form of (cf. \cite{DNW93}) 
\begin{equation}
\omega(x,y)\, =\, \frac{\sigma_{gg}(xys)}{\sigma(s)}
   \, G(x)\, G(y)\, \Theta\left(xy - K^2_{min}\right),
   \label{eq:OMEGA}
\end{equation}
where $G$'s denote the effective number of gluons from the
corresponding projectiles (approximated by the respective gluonic 
structure functions) and $\sigma_{gg}$ and $\sigma$ the gluonic and
hadronic cross sections, respectively.
These are basic formulae of the IGM from which all its 
applications are derived \cite{APPLIC}.\\

\section{Diffractive Dissociation in the IGM approach}

It is now quite straightforward to extend the IGM also to diffractive
dissociation processes. In Fig. 1b we show schematically the IGM 
picture of a diffractive dissociation event. Unlike in Fig. 1a 
here only one of the protons
looses fraction $x$ of its original momentum and gets excited forming a
LJ carrying $x_L= 1 -x$ fraction
of the initial momentum. The other one, which we shall call here the
diffracted proton, looses only a fraction $y$ of its momentum but
otherwise remains intact \cite{FOOT}. In the standard IGM
presented in the previous section we were computing the 
probability $\chi(x,y)$ of
depositing energy fractions $x$ and $y$ in the central region in the
form of gluonic CF of mass
$M = \sqrt{xy\, s}$, which subsequently was decaying and producing
particles. Here we shall be rather interested in the mass $M_X$, a
new variable in our problem, which, as can be seen in 
Fig. 1b, is just the invariant mass of a system composed of the 
CF and the leading jet
formed by one of the colliding protons (we shall call it also
{\it diffractive mass}). Denoting by $E_L$ and $P_L$ the energy and
momentum of the upper (in Fig. 1b ) proton and by $W$ and $P$ the
energy and momentum of the CF,
\begin{equation}
P_L = E_L\, =\, \frac{\sqrt{s}}{2}\, ( 1 - x ), \quad
P   = \frac{\sqrt{s}}{2}\, ( x - y ), \quad
W   = \frac{\sqrt{s}}{2}\, ( x + y ), \label{eq:PLPW}
\end{equation}
the energy $E_X$ and momentum $P_X$ of the diffractive cluster are
given by:  
\begin{equation}
E_X = E_L + W\, =\, \frac{\sqrt{s}}{2}\, (1 + y) \quad {\rm and} \quad
P_X = P_L + P\, =\, \frac{\sqrt{s}}{2}\, (1 - y) \label{eq:EXPX}
\end{equation}
leading to the following expressions for the mass of our diffractive
cluster, $M_X$, and its rapidity, $Y_X$:
\begin{eqnarray}
M_X\, &=&\, \sqrt{E_X^2\, -\, P_X^2}\, \, =\, 
                  \sqrt{s\cdot y}, \label{eq:MASS}\\
Y_X\, &=&\, \frac{1}{2}\, \ln \frac{E_X\, +\, P_X}{E_X\, -\, P_X}\,
          \, =\, \frac{1}{2}\, \ln \frac{1}{y}, \label{eq:RAP}
\end{eqnarray}         
where $\sqrt{s}$ is the invariant energy of the $pp$ system. We are
working in the cm frame of the incoming nucleons. All masses have 
been neglected \cite{FOOT1}.\\

In the limit $y\rightarrow 1$, the whole available energy is stored in
$M_X$ which remains then at rest, i.e., $Y_X = 0$. For small values
of $y$ we have small masses $M_X$ located at large rapidities $Y_X$.
In order to regard our process as being trully of the DD type we 
must assume that all gluons from the target proton participating in 
the collision (i.e., those emitted from the lower vertex in Fig. 1b) 
{\it have to form a colour singlet}. Only then a large rapidity gap 
will form separating the diffracted proton (in the lower part of our 
Fig. 1) and the $M_X$ system (in its upper part), which is the
experimental requirement defining a diffractive event. Otherwise a 
colour string would develop, connecting the diffracted proton and the
diffractive cluster, and would eventually decay, filling the
rapidity gap with produced secondaries. In this way we are
effectively introducing an object resembling closely 
(but by no means identical to) what is known as
Pomeron ($I\!\!P$) and therefore in what follows we shall 
use this notion as a handy abbreviation of notation \cite{FOOTPOM}.
The probability of finding the ``bunch'' of gluons (that forms the
Pomeron) mentioned above in a colour singlet configuration is not
computable at a more fundamental level because the model is simple
and does not involve colour quantum numbers explicitely. However
this probability is represented indirectly in the cross section
between the gluonic bunch and the other proton, the parameter
$\sigma$ discussed below in eq. (\ref{eq:defsig}).

In our approach the definition of the object $I\!\!P$ is essentially 
only kinematical \cite{FOOT2}, very much
in the spirit of those used in all other works which deal with
diffractive processes in the parton and/or string language
\cite{DESAI,INNO,LON,COLL}. We shall therefore try to derive the
whole $M_X^2$ dependence directly from the IGM. 
Our first step
necessary to adapt the standard IGM to DD collisions will be the
introduction a kinematical restriction in the formula for the moments
eq.(\ref{eq:defMOM}) preventing gluons coming from the
diffracted proton (and forming our object $I\!\!P$) to carry more
energy than the one released in the diffractive system. 
Therefore we shall write it as
\begin{eqnarray}
\langle x^ny^m\rangle &=& \int_0^1\! dx\,x^n\, \int_0^{y_{max}}\! dy\, 
y^m\, \omega (x,y), \label{eq:defMOMD}
\end{eqnarray}
where $y_{max} = \frac{M^2_X}{s}$ and the meaning of the spectral function
$\omega(x,y)$ remains the same as before.\\

We can now calculate the diffractive mass distribution $M_X$ using
the $\chi(x,y)$ function by simply performing a change of variables (cf.
eq.(\ref{eq:MASS})),  
\begin{eqnarray}
\frac{dN}{dM_X^2}\, &=&\, \int_0^1\! dx\, \int_0^1\! dy\, 
                       \chi(x,y)\, \delta\left( M^2_X - sy\right)\, 
                       \Theta \left( xy - K^2_{min}\right)
                       \nonumber \\
                    &=& 
          \frac{1}{s}\, \int_{_{\frac{m_0^2 }{M_X^2}}}^1\! dx\,
          \chi\left( x,\frac{M_X^2}{s}\right), \label{eq:RES}
\end{eqnarray}                       
In the IGM \cite{IGM,DNW93} the distribution $\chi(x,y)$ is a wide
gaussian in the variables $x$ and $y$ changing slowly with the energy
$\sqrt{s}$. Substituting now eq.(\ref{eq:CHI}) into eq.(\ref{eq:RES})
we arrive at the following simple  expression for the diffractive
mass distribution:  
\begin{equation}
\frac{dN}{dM^2_X}\, =\, \frac{1}{s}\, \cdot\, F(M^2_X,s) 
                   \cdot H(M_X^2,s) \label{eq:DIS}
\end{equation}           
where 
\begin{eqnarray}
F(M^2_X,s) &=&  \exp \left[ - \frac{\langle x^2\rangle}{2D_{xy}}\,
                 \left( \frac{M^2_X}{s} - \langle y\rangle \right) ^2
                 \right] \label{eq:FFF}
\end{eqnarray}
and
\begin{eqnarray}
H(M^2_X,s) &=& \frac{\chi_0}{2\pi \sqrt{D_{xy}}}\, 
  \int^1_{\!_{\frac{m_0^2 }{M_X^2}}}\! dx
         \cdot \label{eq:H}\\
      && \cdot \exp \left\{ - \frac{1}{2D_{xy}}\,
               \left[ \langle y^2\rangle (x - \langle x\rangle )^2 -
               2\langle xy\rangle (x - \langle x\rangle) 
               \left( \frac{M_X^2}{s} - \langle y \rangle \right)
               \right] \right\} .\nonumber 
\end{eqnarray}
The moments  $\langle q^n\rangle,~~q=x,y$ and $n=1,2$ are given by
(\ref{eq:defMOMD}) and are the only place where dynamical quantities
like the gluonic and hadronic cross sections appear in the IGM. At this
point we emphasize that we are all the time dealing with a proton-
proton scattering. However, as was said above, we are in fact
selecting a special class of events and therefore we must choose the
correct dynamical inputs in the present situation, specially the
gluon distribution inside the diffracted proton and the hadronic
cross section $\sigma$ appearing in $\omega$. As a first
approximation we shall take $G^{\rm I\!\!P}(y) = G^p(y) = G(y)$ (cf.
\cite{IS}), with $G(x) = p(m+1)\frac{(1-y)^m}{y}$, with m=5, the same
already used by us before \cite{DNW93}. The amount of
the diffracted nucleon momentum, {\it p}, allocated specifically to the
$I\!\!P$ gluonic cluster and the hadronic cross section $\sigma$ are
both unknown. However, they always appear as a ratio
($\frac{p}{\sigma}$) of parameters in $\omega$ and different choices
are possible. Just in order to make contact with the present
knowledge about the Pomeron, we shall choose
\begin{eqnarray}
\sigma(s) \,&=& \, \sigma^{\rm I\!\!P p} \, = \, a + 
b \, \ln \frac{s}{s_0}
\label{eq:defsig}
\end{eqnarray}
where $s_0 = 1$ $GeV^2$ and a and b are parameters 
to be fixed from data
analysis. As it will be seen, $\sigma(s)$ turns out to be a very
slowly varying function of $\sqrt{s}$ assuming values between 2.6
and 3.0 mb, which is a well accepted value for the
Pomeron-proton cross section, and $ p \simeq 0.05 $.\\ 

Before performing a full numerical calculation let us estimate
eqs.(\ref{eq:DIS}) and (\ref{eq:H}) keeping only the most singular
parts of the gluonic distributions used (i.e., $G(x)\simeq 1/x$) and
collecting all other factors in eq.(\ref{eq:OMEGA}) in a single
parameter $c$. Let us first assume that the ratio of the cross
sections $\frac{\sigma(xys)}{\sigma(s)}$ does not depend on $x$ and
$y$. Neglecting all terms of the order of $\frac{m_0^2}{s}$ and
$\frac{m_0^2}{M_X^2}$ we arrive at the following expressions for the
moments calculated in eq.(\ref{eq:defMOMD}): 
\begin{eqnarray}
\langle x\rangle \, &=& 2\, \langle x^2\rangle \simeq\,
                    c\cdot \ln \frac{M_X^2}{m_0^2};  \label{eq:momX}\\
\langle y\rangle \, &=& 2\, \frac{s}{M_X^2}\, \langle y^2\rangle
                       \simeq \, c\cdot \frac{M_X^2}{s}\cdot 
                    \ln \frac{M_X^2}{m_0^2}; \label{eq:newY}\\
\langle x\cdot y\rangle \, &\simeq& \, c\, \left(\frac{M_X^2}{s}\, -
           \frac{m_0^2}{s}\cdot \ln\frac{M_X^2}{m_0^2}\right).
            \label{eq:momXY}
\end{eqnarray}           
Notice that in all cases of interest $\langle x\cdot y\rangle$ is
much smaller than other moments (by a factor
$\ln\frac{M_X^2}{m_0^2}$, at least). It means that $D_{xy}\, \simeq\,
\langle x^2\rangle\langle y^2\rangle $ and consequently
\begin{eqnarray}
F(M^2_X,s)\, &\simeq&\, \exp\left[ - \frac{\left(\frac{M_X^2}{s}\, -\, 
                       \langle y\rangle\right)^2}
                       {2\, \langle y^2\rangle}\right] \nonumber\\
	     &\simeq&\, \exp\left[ - \frac{\left(1\, -\, 
                        c\cdot \ln\frac{M_X^2}{m_0^2}\right)^2}
                        {c\cdot \ln\frac{M_X^2}{m_0^2}}\right]
\label{eq:F}
\end{eqnarray}
and 
\begin{eqnarray}
H(M_X^2,s)\, &\simeq&\, \frac{\chi_0}{2\pi \sqrt{D_{xy}}}\, 
                  \int^1_{\!_\frac{m_0^2}{M_X^2}}\! dx
                   \exp\left[ - \frac{(x - \langle x\rangle
                    )^2}{2\langle x^2\rangle}\right] 
            \simeq\, {\rm const}\, \cdot\, 
            \frac{\sqrt{\langle x^2\rangle}}{\sqrt{D_{xy}}} 
            \, =\, {\rm const}\, \cdot \, 
            \frac{1}{\sqrt{\langle y^2\rangle}}\nonumber\\
             &\simeq&\, {\rm const}\cdot \frac{s}{M_X^2
                \cdot \sqrt{c\cdot \ln\frac{M_X^2}{m_0^2}}} \label{eq:newapprH}
\end{eqnarray}
leading to
\begin{eqnarray}
\frac{dN}{dM_X^2}\, &\simeq&\, \frac{1}{s}\, \cdot\, H(M_X^2,s)\cdot
			       F(M^2_X,s) \nonumber\\
              &\simeq&\, \frac{\rm const}{M_X^2}\cdot 
                        \frac{1}{\sqrt{c\cdot \ln\frac{M_X^2}{m_0^2}}} \cdot 
                        \exp\left[ - \frac{\left(1\, -\, 
                        c\cdot \ln\frac{M_X^2}{m_0^2}\right)^2}
                        {c\cdot \ln\frac{M_X^2}{m_0^2}}\right].
                        \label{eq:result}
\end{eqnarray}                        

The expression above is governed by the $\frac{1}{M^2_X}$ term. The
other two terms have a weaker dependence on $M^2_X$. They distort the
main ($\frac{1}{M^2_X}$) curve in opposite directions and tend to 
compensate each other. It is therefore very interesting to note
that even before choosing a very detailed form for the gluon 
distributions and hadronic cross sections we obtain analytically the
typical shape of a diffractive spectrum.

\section{Comparison with experimental data}
 In Fig. 2a we show our diffractive mass spectrum and compare 
it to experimental data from the CERN-ISR \cite{ISDATA}, which are
usually parametrized by the form $\frac{1}{M^2_X}$. These spectra were 
calculated with expression (\ref{eq:DIS}) with the same gluon 
distributions, cross sections and the mass parameter $m_0$ 
($m_0 = 350 $ MeV) used in previous works \cite{IGM,DNW94,DNW93}.
As it can be seen, the agreement between our curves and data is
reasonable. At large values of $\frac{M^2_X}{s}$, experimental points
start to flatten out, deviating from the $\frac{1}{M^2_X}$ behaviour.
This may be due to the contribution of non-diffractive events. We
expect therefore some discrepancy between theory and experiment in
this region. At very low values of $\frac{M^2_X}{s}$ and lower
energies $\sqrt{s}$ our model does not give a good description of
data. Here again some discrepancy should be expected because we
are neglecting all resonance effects. In principle a better agreement
between theory and data could be achieved in this region. It would
be enough to choose $ m_0 = 550 $ MeV keeping everything else as
before. The result is shown in Fig. 2b. A discrepancy persists at
very low masses at the lowest center of mass energy $\sqrt{s}$. 
However this region is almost beyond the validity domain of the model. 
Instead of changing $m_0$ (which would also require a change
in our previous works), we prefer to keep its usual value. Given the
qualitative nature of the present work, we choose to be consistent with
our previous works in prejudice of the quality of the fits. 

Fig.3 and 4 show similar comparisons for $\sqrt{s}=
546$ and $ 1800  $ GeV respectively. Data are from refs. \cite{Bozzo}
and \cite{DATA}. Again we find reasonable agreement with experiment.
In the low $\frac{M^2_X}{s}$ region at higher $\sqrt{s}$ the
agreement is very good. All the curves above were obtained with
$a = 2.6 $~mb and $b = 0.01$~mb. \\

As a straightforward extension of our calculation we now apply
expression (\ref{eq:DIS}) to the study of diffractive pion-proton and
kaon-proton scattering. We first consider the cases 
$ p+\pi \rightarrow p + X$ and $p + K \rightarrow p +X$.
 This corresponds to replace the proton by a
pion or a kaon in the upper line of Fig. 1b, everything else remaining
the same. We must also substitute the gluon distributions in the
proton, $G(x)$, by the corresponding gluon distributions in the pion
and kaon, taken from
\cite{MRS}. Here, for simplicity, we take $ G^{\pi}(x) = G^{K}(x)$.
This is supported by an ACCMOR collaboration data analysis
\cite{ACC}. We also assume that $\sigma^{\rm I\!\!P p} = \sigma^{\rm
I\!\!P \pi} = \sigma^{\rm I\!\!P K} $. The comparison between our
results and data from the EHS/NA22 collaboration \cite{EHS} is shown
in Figs. 5a (pions) and 5b (kaons). 
We may also have diffracted mesons, which undergo
reactions of the type $\pi +p \rightarrow \pi + X$ and 
$K + p \rightarrow K + X$. This corresponds to replace the proton by
a pion or a kaon in the lower line of Fig. 1b, substituting also the
corresponding gluon distributions. The comparison between our results
and experimental data \cite{EHS} is shown in Fig. 6a (pions) and 
6b (kaons). As it can be seen, a good description of data is
obtained. \\

Let us consider now the energy dependence of our results.
The CDF collaboration studied single diffractive events and found
some energy dependence in the diffractive mass spectrum.  This fact
is illustrated by writing 
\begin{equation}
\frac{s}{\sigma_{SD}}\, \frac{d\sigma_{SD}}{dM^2_X} \propto
           \frac{1}{\left( M^2_X\right) ^{1+\epsilon}} \label{eq:ED}
\end{equation}
where the factor $\epsilon$ which has been reported to be \cite{GOUL}
$\epsilon = 0.121\pm 0.011$ at $\sqrt{s} = 546$ GeV and $\epsilon =
0.103\pm 0.017$ at $\sqrt{s} = 1800$ GeV, respectively. Considering
the error bars one might say that this value is just constant (and
that there would be no indication of energy dependence), but a real
(albeit small one) change in $\epsilon$ is not excluded. Therefore,
if confirmed, it would mean that the distribution becomes slightly
broader.\\ 

In the IGM everything is from the beginning energy dependent and so
should be the diffractive mass distribution. We start analysing the
analytical approximation eqs. (\ref{eq:DIS},\ref{eq:F},
\ref{eq:newapprH}). In (\ref{eq:newapprH}) we see that the $s$-dependence
factorizes and the function $H(M^2_X)$ has the same shape for all energies, 
the difference being only a multiplying factor (in numerical calculations
this behaviour is slightly violated). In (\ref{eq:F}) the
$s$-dependence does not factorize and remains in the moments or, 
equivalently,
in the variable $c$ . $F(M^2_X)$ is a broad  function with 
maximum value determined by the moment~$\langle y \rangle $ which 
increases with the
energy, making $F$ to "rotate" in a way that it becomes higher at 
lower values
of $M^2_X$ and becomes deeper at larger values of $M^2_X$. When
we multiply $H$ (which goes essntially like $1 / M^2_X$) by $F$ it becomes
steeper. This behaviour of $F$ and $H$ is illustrated in Fig. 7a and 7b 
respectively, which show the numerical evaluation of eqs. (\ref{eq:FFF})
and (\ref{eq:H}).
The result of the numerical evaluation of eq. (\ref{eq:DIS})
is presented in Fig. 8. There we show the energy
dependence of our diffractive mass spectra in proton-proton
scattering. Fig. 8a shows diffractive mass spectra for $\sqrt{s}=
23.5$ GeV (solid lines), $44.6$ GeV (dashed lines) and $62.4$ GeV
(dotted lines). Fig. 8b shows spectra at $\sqrt{s}=0.54$ TeV (solid
lines), $0.9$ TeV (dashed lines) and $1.8$ TeV (dotted lines). Finally
Fig. 8c shows our prediction for the diffractive mass spectrum at
LHC ($\sqrt{s}= 14 $ TeV) compared to the Tevatron one. The spectra
in Fig. 8a and 8b are the same of Figs. 2, 3 and 4. It is interesting to
note that the behaviour that we find is not in contradiction with
data. In all curves we observe a modest 
narrowing as the energy increases. 
This
small effect means that the diffractive mass becomes a smaller fraction
of the available energy $\sqrt{s}$. In other words, the "diffractive
inelasticity" decreases with energy and consequently the "diffracted
leading particles" follow a harder $x_F$ spectrum. Physically, in
the context of the IGM, this means that the deposited energy is
increasing with $\sqrt{s}$ but it will be mostly released outside the 
phase space region that we are selecting. A measure of the
"diffractive inelasticity" is the quantity $\xi=\frac{M^2_X}{s}$.
Making a trivial
change of variables in eq.(\ref{eq:DIS}) we can calculate its 
average value $\langle \xi \rangle$ :
\begin{eqnarray}
\langle \xi \rangle \left( s \right) &=& \int_{\xi_{min}}^{\xi_{max}}
d\xi\, \frac{d N}{d \xi}\,  \xi \label{eq:XI}
\end{eqnarray}                       
where $\xi_{min}$ ($=\frac{1.5}{s}$) and $\xi_{max}$ ($= 0.1$) are the
same used in (\cite{GOUL}) for the purpose of comparison.
In Fig. 9 we
plot $\langle \xi \rangle $ against $\sqrt{s}$. As it can be seen
$\langle \xi \rangle$ decreases with $\sqrt{s}$ not only because
$\xi_{min}$ becomes smaller but also because $\frac{d N}{d \xi}$
changes with the energy, falling faster. This qualitative 
behaviour of
$\langle \xi \rangle$ is in agreement with the estimate of the
same quantity extracted from cosmic ray data analysis \cite{COV}.
Also shown in Fig. 9 is the quantity 
$\langle \xi^{\varepsilon} \rangle$ (which
has been discussed in \cite{GOUL} in connection with the energy
dependence of the single diffractive cross-section) for 
$\varepsilon= 0.08$~(dashed lines) and~$\varepsilon= 0.112$
~(dotted lines).\\

The energy behaviour of $\frac{d N}{d M^2_X}$ is determined by the
moments (\ref{eq:defMOMD}) and (\ref{eq:OMEGA}). 
Since $\sigma_{gg}(s)$ and $G(x)$ are the same as in 
previous works, being thus fixed, the only source of uncertainty in
the $s$-dependence of the results is in the ratio $\frac{p}{\sigma}$, 
which is the only 
free parameter in the model. All curves presented above were obtained 
with the 
choice (\ref{eq:defsig}) made for $\sigma$ and with $p \simeq 0.05$.
We have checked that
for a stronger growth of $\sigma^{\rm I\!\!P p}$  with $\sqrt{s}$ the
energy
behaviour of $\frac{d N}{d M^2_X}$ might become even the opposite of the
one found here, i.e., the diffractive mass distribution would become
broader at higher energies. However these strongly $s$-dependent 
parametrizations do not give an acceptable description of  the
existing data and were therefore excluded. Considering what was
said above one might think that we can discriminate between
different Pomeron-proton cross sections and we could use this model
to extract $\sigma^{\rm I\!\!P p}$ from data. We stress however that, in
this model, only the ratio $\frac{p}{\sigma}$ enters effectively in
the calculations and it is impossible to completely disentangle
these two variables. In this sense, the almost constancy (with
$\sqrt{s}$) of $\sigma^{\rm I\!\!P p}$  may be just an indication
of some increase of $p$ with $\sqrt{s}$ in a way that the ratio
$\frac{p}{\sigma}$ "scales " with the energy.\\ 

\section{Conclusions}
To conclude, we have shown that:
\begin{itemize}
\item[$(i)$] The original IGM
(\cite{IGM,DNW94,DNW93}) is reasonably successful in describing
non-diffractive events. With only two natural changes, namely
the introduction of the kinematical cut-off 
$y_{max}=\frac{M_X^2}{s}$ and
the multiplication of the function $\omega$ by a constant factor
reflecting the essentially unknown combination of $I\!\!P$-proton
cross section $\sigma$ and the amount of gluonic energy-momentum of
the diffractive proton allocated to the object $I\!\!P$, $p$ (in form of
the ratio $\frac{p}{\sigma}$), (with all other parameters kept as in 
previous applications of IGM) it turns out to be also able to provide
a reasonable description of diffractive events \cite{FOOT4} and their
energy dependence. We predict that, at higher energies, a narrowing
of the $M^2_X$ distribution may be observed.

\item[$(ii)$] As is obvious from Fig. 1b, it provides a detailed
description of the diffractive cluster (of mass $M_X$), {\it
disentangling} it in a natural way into gluonic cluster of mass $M$
containing the majority of produced secondaries (which corresponds to
CF in the original IGM) and leading jet (inside the diffractive system)
with momentum fraction $x_L$ carrying quantum numbers of the
diffractively excited projectile. This fact may be very usefull for
the cosmic ray applications of the IGM, in particular to studies
using DD like those presented in \cite{CAP,COSMIC} (where such
disentanglement seems to be important and so far was introduced in an
{\it ad hoc} way only).
\end{itemize}

\vspace{0.5cm}
\underline{Acknowledgements}: This work has been supported by FAPESP,
CNPQ (Brazil) and KBN (Poland). F.S.N. is deeply indepted to his Polish
collegaues from SINS, Warsaw, for the hospitality extended to
him during his stay there. We would like to warmly thank R. Covolan and
Y. Hama for many fruitful discussions.


\vspace{1cm}
\noindent
{\bf Figure Captions}\\
\begin{itemize}
\item[{\bf Fig. 1}] IGM description of a proton-proton scattering 
: a) general case. b) with the formation of a diffractive 
system of invariant mass $M_X$.

\item[{\bf Fig. 2}] a) Diffractive mass spectrum for $pp$
collisions calculted with the IGM (eq.(\ref{eq:RES})) and compared
with CERN-ISR data \cite{ISDATA}. b) the same as a) with
$m_0 = 550$ MeV. 

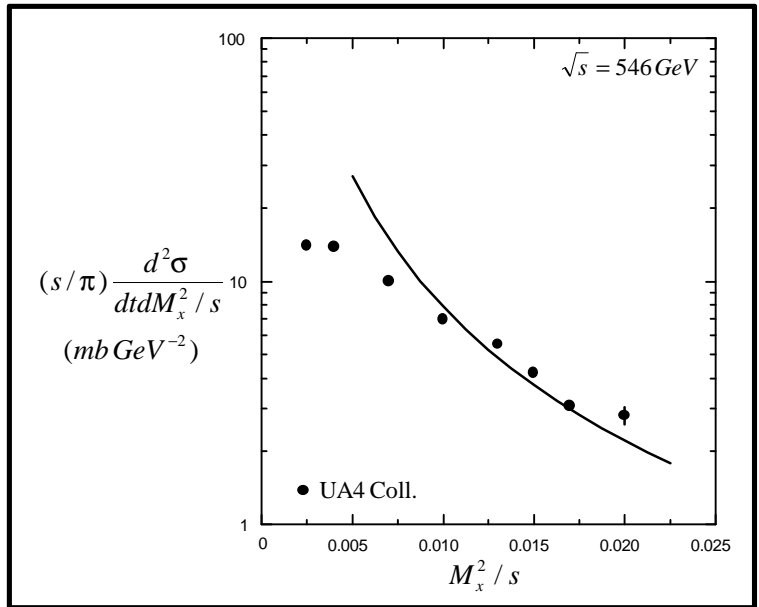
\item[{\bf Fig. 3}] Diffractive mass spectrum for $p\bar{p}$
collisions calculted with the IGM (eq.(\ref{eq:RES})) and compared
with CERN-SPS Collider data \cite{Bozzo}. 
  
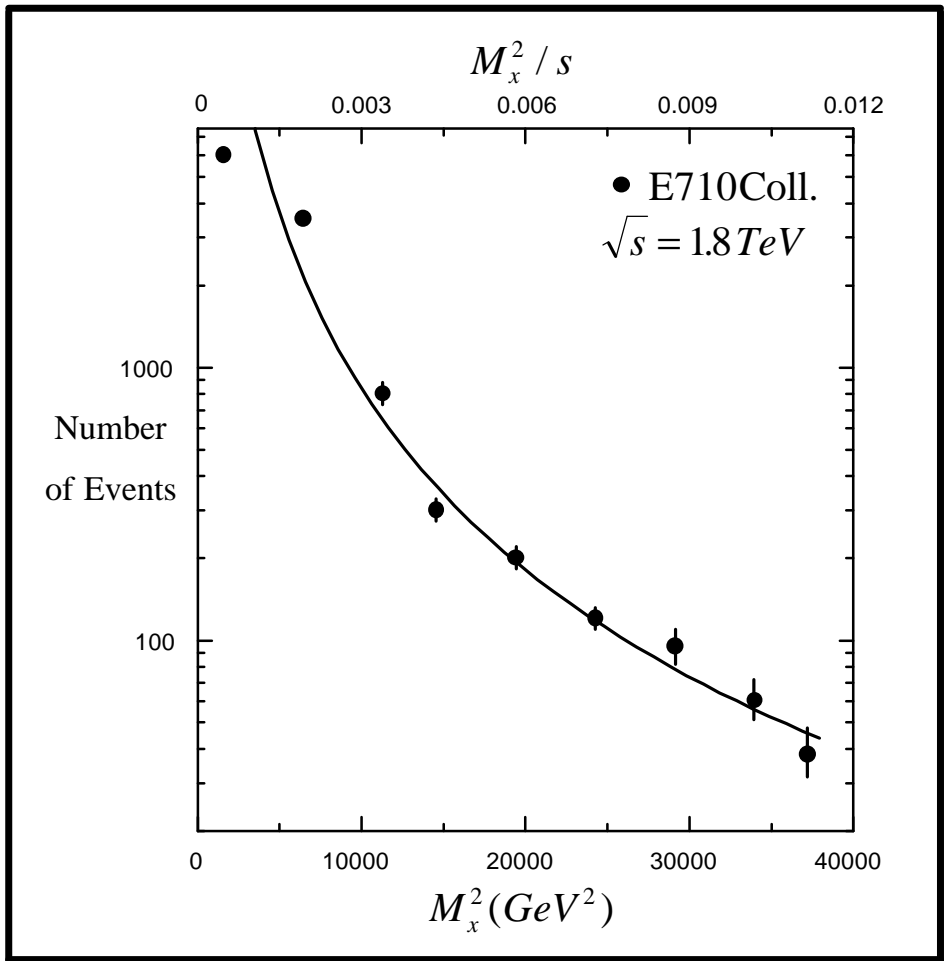
\item[{\bf Fig. 4}] Diffractive mass spectrum for $p\bar{p}$
collisions calculted with the IGM (eq.(\ref{eq:RES})) and compared
with FERMILAB Tevatron data \cite{DATA}. 

\item[{\bf Fig. 5}] a) Diffractive mass spectrum for 
$p + \pi^+ \rightarrow p +X $
collisions calculted with the IGM (eq.(\ref{eq:RES})) and compared
with experimental data \cite{EHS}. b) The same as a) for
$p + K^+ \rightarrow p +X $ collisions. 

\item[{\bf Fig. 6}] a) Diffractive mass spectrum for 
$\pi^+ + p \rightarrow \pi^+ + X$ collisions
calculted with the IGM (eq.(\ref{eq:RES}) and compared with
experimental data \cite{EHS}. b) The same as a) for
$K^+ + p \rightarrow K^+ + X$ collisions.

\item[{\bf Fig. 7}] a) Energy dependence of the function 
$F \left( M^2_X,s \right)$ (\ref{eq:FFF}).
b) Energy dependence of the function
$H \left( M^2_X,s \right)$ (\ref{eq:H})
calculted with the IGM.
 
\item[{\bf Fig. 8}] a) Energy dependence of diffractive mass spectra
calculated with the IGM (eq. (\ref{eq:DIS})).
The solid, dashed and dotted lines represent
spectra at $\sqrt{s}= 23.5$, $44.6$ and $62.4$ GeV respectively. b)
The same as a) for $\sqrt{s}= 0.54$ (solid lines), $0.90$ (dashed lines)
and $1.8$ TeV (dotted lines). c) The same as a) for $\sqrt{s}= 1.8$
(solid lines) and $14$ TeV (dashed lines).

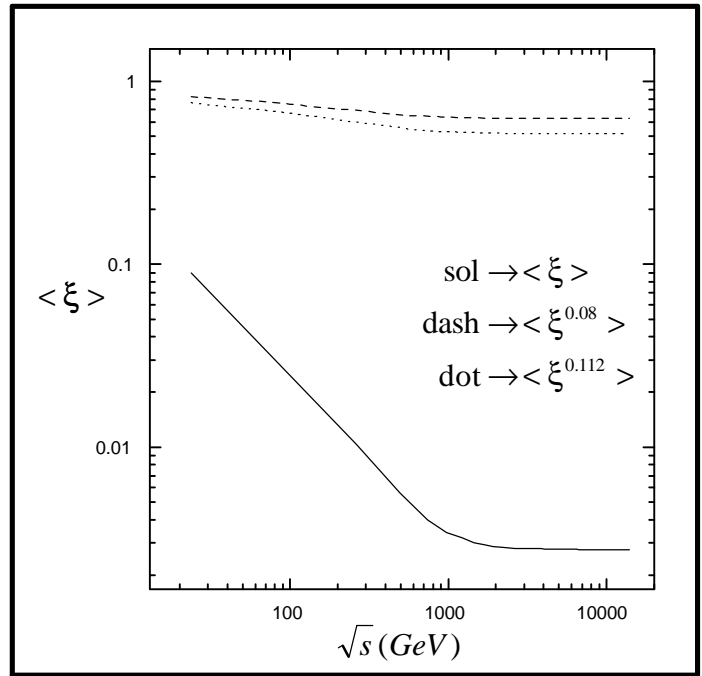
\item[{\bf Fig. 9}] Energy dependence of the "diffractive inelasticity"
$\langle \xi \rangle$ and of $\langle \xi^{\varepsilon} \rangle$.

\end{itemize}


\begin{thebibliography}{99}
\bibitem{FOOT} It can be also deflected with some invariant momentum
               transfer $t$. However, because the IGM does not 
               include so far any transverse momentum transfers,
               the results presented here must be regarded as
               appropriately averaged/integrated over $t$.
\bibitem{FOOT1} As everywhere in the IGM (\cite{IGM,DNW94,DNW93}) all
                masses are neglected here. This implies that our 
                approach will be the more accurate the higher is the  
                energy of the reaction and also that it will not be 
                valid for very small values of $M^2_X$. The IGM is 
                applicable (and was tested) from
                energies $\sqrt{s} \geq 10$ GeV onward only and this 
                value imposes a lower limit of applicability for the
                model.
\bibitem{FOOTPOM} Since already some time \cite{IS} the Pomeron 
                  is treated as being
                  composed of partons, i.e., gluons and sea 
                  $\bar{q}q$ pairs, in much
                  the same way as hadrons, with some 
                  characteristic distribution
                  functions which will presumably become 
                  accurately known in HERA
                  experiments \cite{zeus}.
\end{thebibliography}
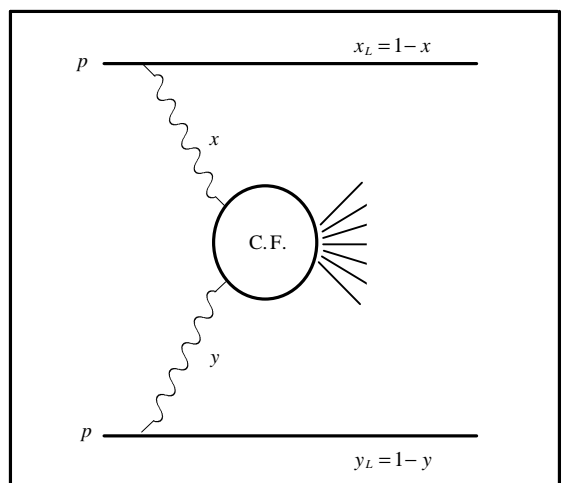
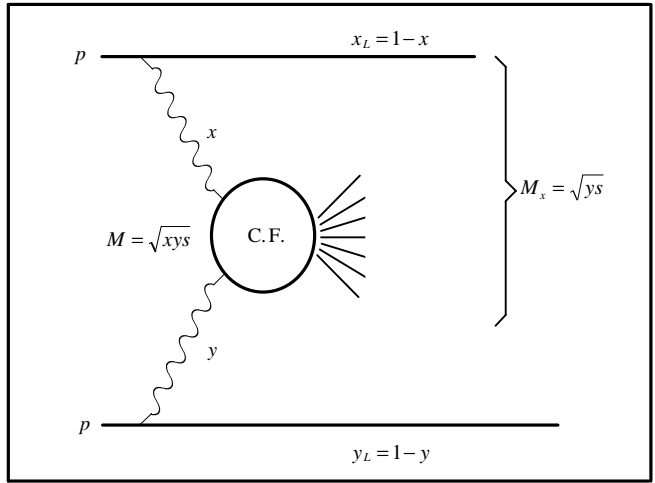
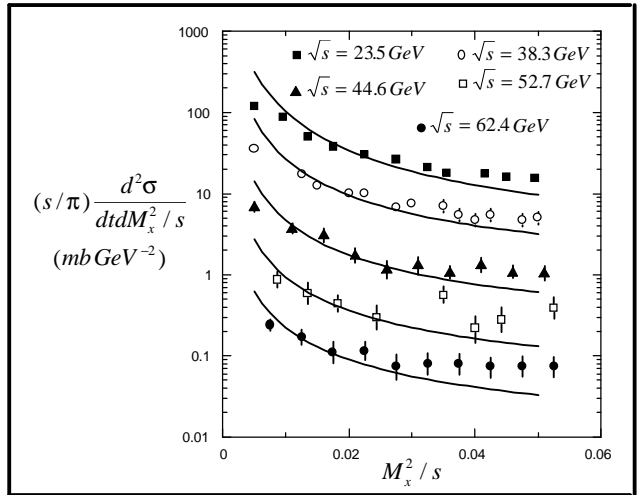
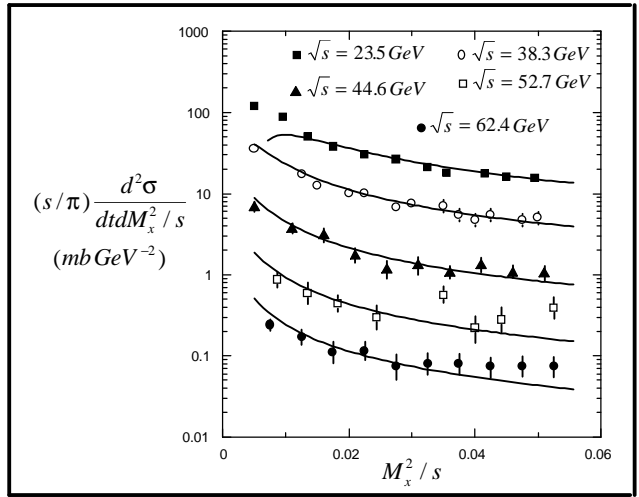
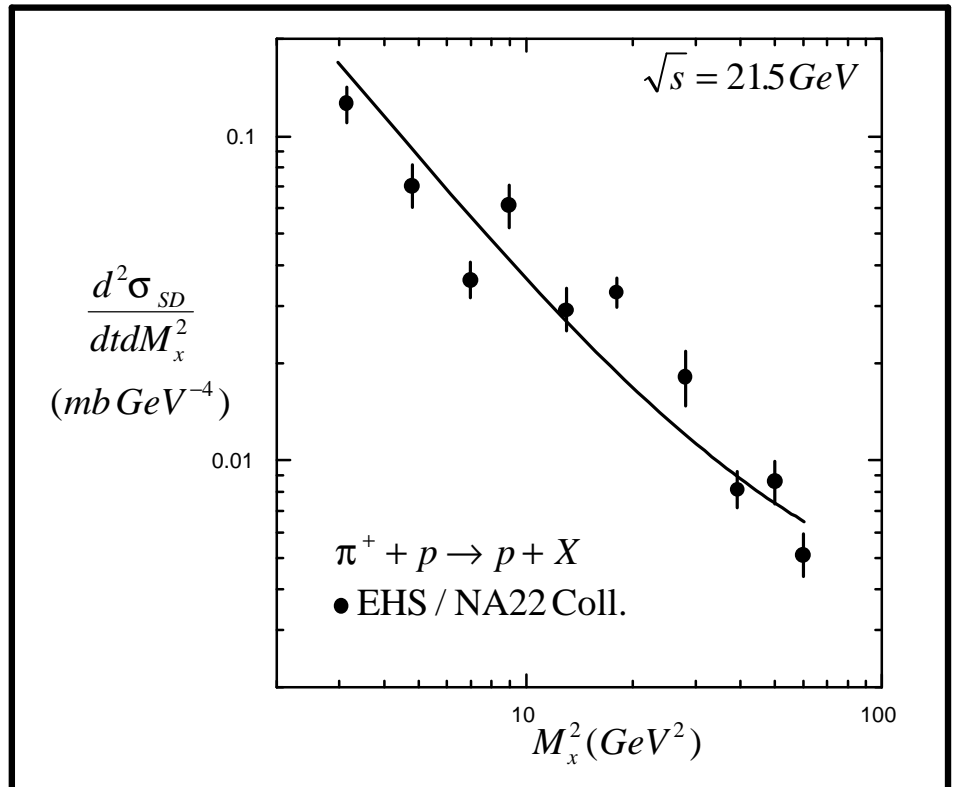
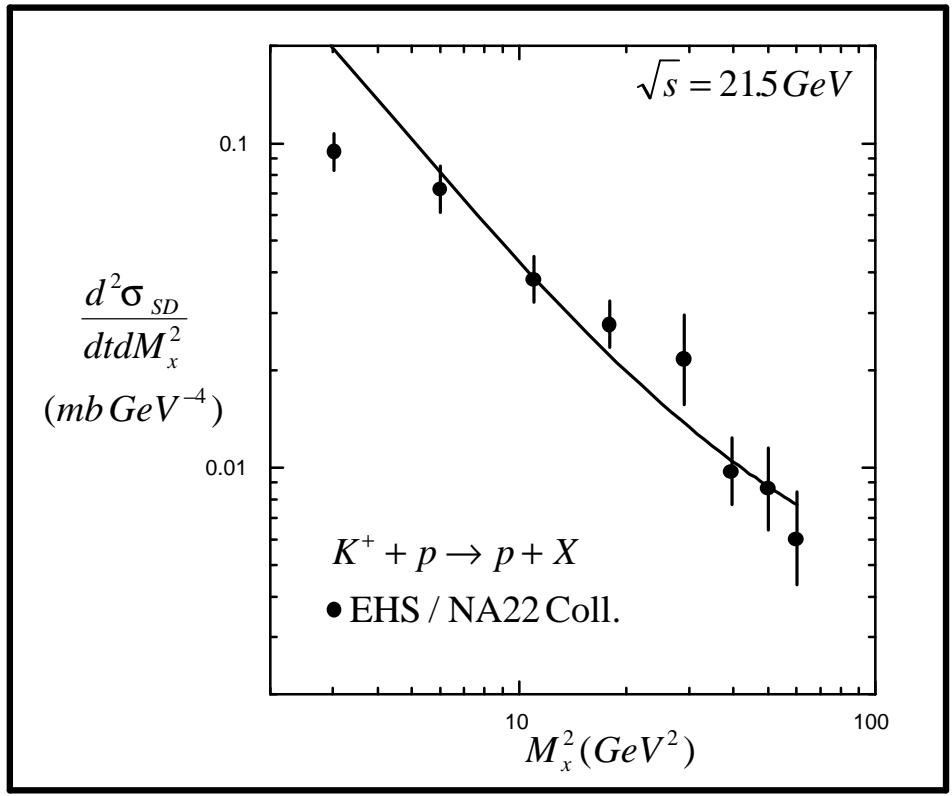
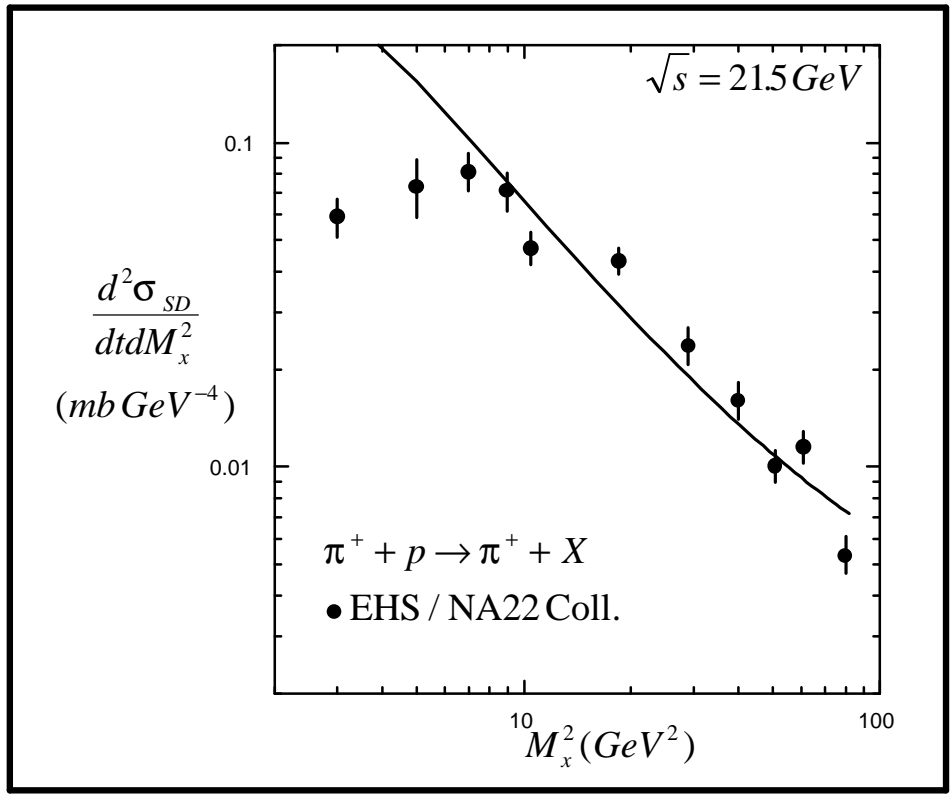
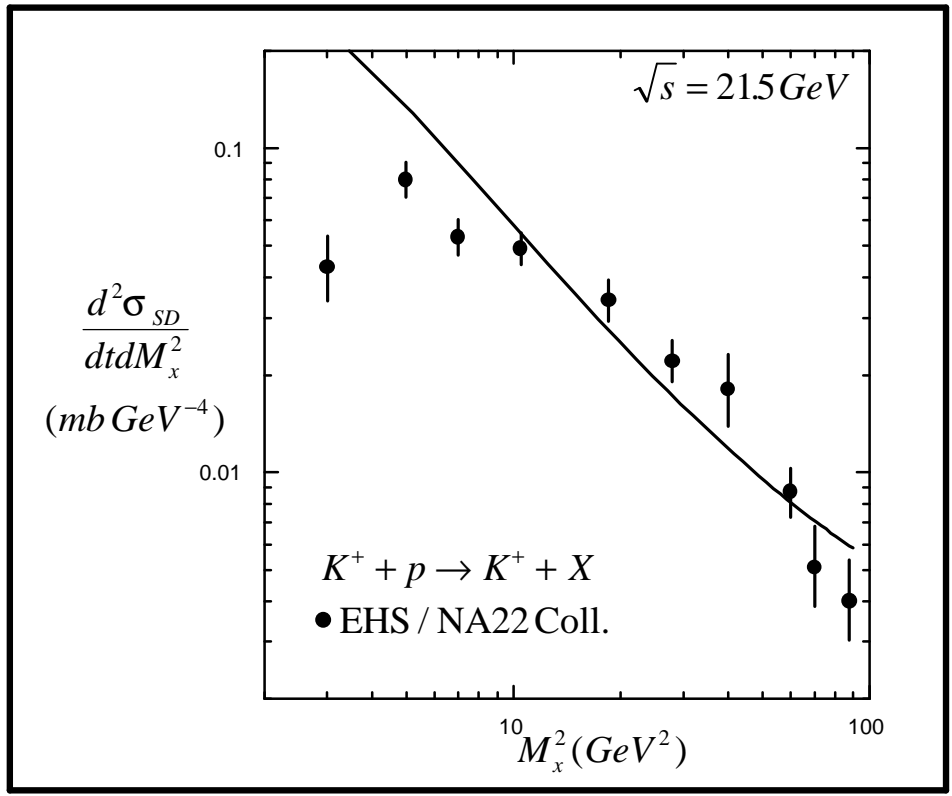
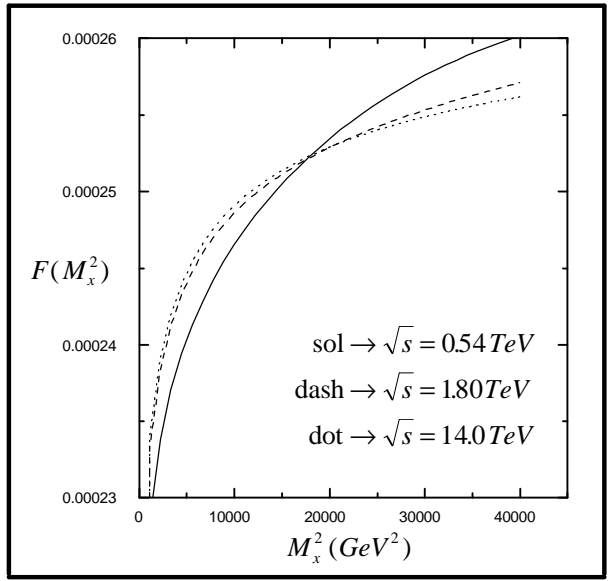
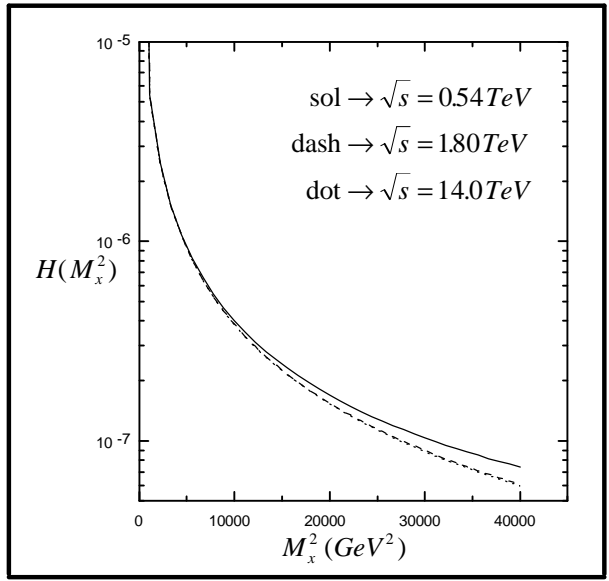
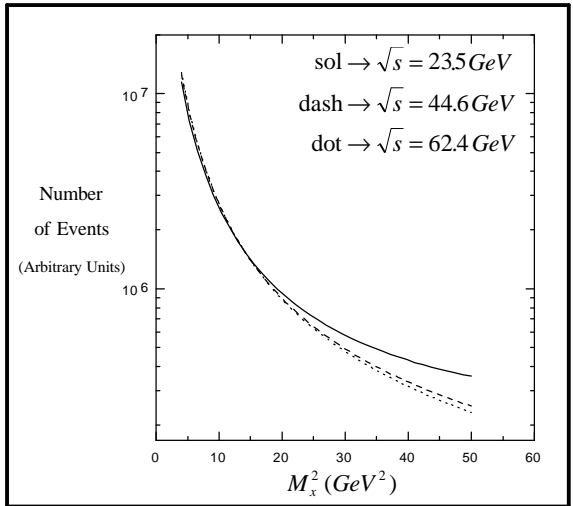
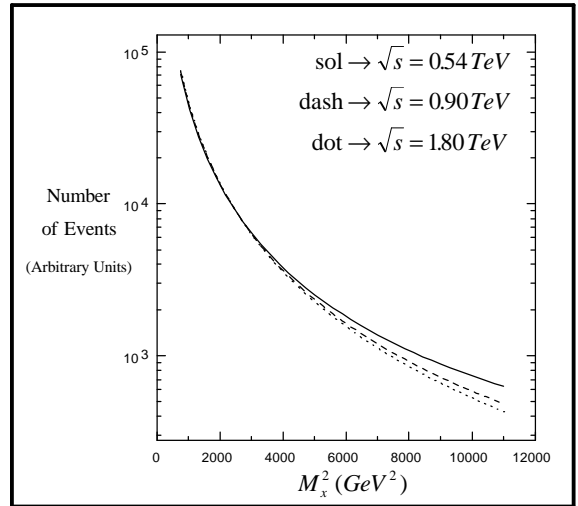
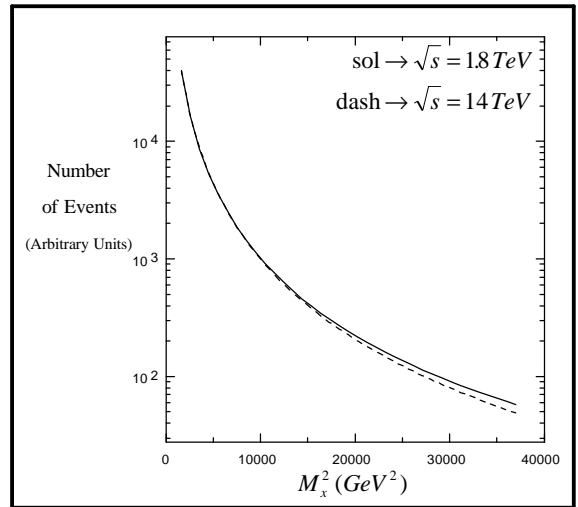
\end{document}